\documentclass[11pt]{article}
\usepackage{graphicx}
\usepackage{amsmath}
\usepackage{amssymb}
\usepackage{amsxtra}
\def\Journal#1#2#3#4{{#1} {\bf #2} (#4) #3 }

\def\PLB{{ Phys. Lett.}  B}

\def\PRD{{ Phys. Rev.} D}

\def\GaC{ Gravitation and Cosmology}

\def\JETPL{ JETP Lett.}

\def\CQG{ Class. Quantum Grav.}

\def\MPLA{{ Mod. Phys. Lett.}  A}
\def\IJTP{ Int. J. Theor. Phys.}
\def\IJMPA{{ Int. J. Mod. Phys.}  A}

\def\NJP{ New J. of Phys.}

\def\BWP{ Bled Workshops in Physics}
\def\s{{\,\rm s}}

\def\GeV{\,{\rm GeV}}

\def\({\left(}
\def\){\right)}

\title{Dark Atoms and their decaying constituents}
\author{K.~Belotsky$^{1,2}$, M.~Khlopov$^{1,2,3}$, M.~Laletin$^{1}$\\$^{1}$National Research Nuclear University "MEPHI"\\(Moscow Engineering Physics Institute),\\ 115409 Moscow, Russia \\$^{2}$ Centre for Cosmoparticle Physics ``Cosmion"\\ 115409 Moscow, Russia \\$^{3}$ APC laboratory 10, rue Alice Domon et L\'eonie Duquet \\75205 Paris Cedex 13, France\\}
\date{}

\begin{document}
\maketitle
\begin{center}

\end{center}

\begin{abstract}
The nonbaryonic dark matter of the Universe might consist of
new stable charged species, bound by ordinary Coulomb
interactions in various forms of heavy neutral "dark atoms". The existing models offer natural implementations for the dominant and subdominant forms of dark atom components. In the framework of Walking Technicolor the charge asymmetric excess of both stable negatively doubly charged technilepton $\zeta^{--}$ and metastable but longliving positively doubly charged technibaryon $UU^{++}$ can be generated in the early Universe together with the observed baryon asymmetry.
If the excess of $\zeta$ exceeds by several orders of magnitude the excess of  $UU$, dark matter might consist dominantly by $He\zeta$ dark atoms of nuclear interacting O-helium ($OHe$)  bound state of $\zeta$ with primordial helium.
This dominant dark matter component causes negligible nuclear recoil in underground experiments, but can explain positive results of DAMA/NaI and DAMA/LIBRA experiments by annual modulations of few keV energy release in radiative capture of OHe by sodium.
However, a sufficiently small subdominant component of WIMP-like objects $UU\zeta$  can also form. Making up a small fraction of dark matter, it can also evade the severe constraints on WIMPs from underground detectors. Although sparse, this subdominant component can lead to observable effects, since leptonic decays of technibaryons $UU$ give rise to two positively charged leptons  contrary to the pairs of opposite charge leptons created in decays of neutral particles. We show that decays of $UU^{++}\rightarrow e^+ e^+, \mu^+ \mu^+, \tau^+ \tau^+$ of the subdominant $UU\zeta$ component of dark matter,  can explain the observed high energy positron excess in the cosmic rays if the fraction of $UU\zeta$ is $\sim 10^{-6}$ of the total dark matter density, the mass of $UU^{++}$ about 1 TeV and the lifetime about $10^{20} \s$. Optimizing fit of recent AMS-02 data by model parameters, the predicted mass range of such long-living double charge particle is challenging for its search at the LHC.
\end{abstract}

\section{Introduction}\label{intro}
Dark atoms offer an interesting possibility to solve the puzzles of dark matter searches. It turns out that even stable electrically charged particles can exist hidden in such atoms,  bound by  ordinary Coulomb interactions (see \cite{mpla,DMRev,DDMRev} and references therein).
Stable particles with charge -1 are excluded due to overproduction of anomalous isotopes.  However, there doesn't appear such an evident contradiction for negatively doubly charged particles.

There
exist several types of particle models where heavy
stable -2  charged species, $O^{--}$, are predicted:
\begin{itemize}
\item[(a)] AC-leptons, predicted
as an extension of the Standard Model, based on the approach
of almost-commutative geometry \cite{Khlopov:2006dk,5,FKS,bookAC}.
\item[(b)] Technileptons and
anti-technibaryons in the framework of Walking Technicolor
(WTC) \cite{KK,Sannino:2004qp,Hong:2004td,Dietrich:2005jn,Dietrich:2005wk,Gudnason:2006ug,Gudnason:2006yj}.
\item[(c)] stable "heavy quark clusters" $\bar U \bar U \bar U$ formed by anti-$U$ quark of 4th generation
\cite{Khlopov:2006dk,Q,I,lom,KPS06,Belotsky:2008se} \item[(d)] and, finally, stable charged
clusters $\bar u_5 \bar u_5 \bar u_5$ of (anti)quarks $\bar u_5$ of
5th family can follow from the approach, unifying spins and charges\cite{Norma}.
\end{itemize}
All these models also
predict corresponding +2 charge particles. If these positively charged particles remain free in the early Universe,
they can recombine with ordinary electrons in anomalous helium, which is strongly constrained in
terrestrial matter. Therefore a cosmological scenario should provide a  mechanism which suppresses anomalous helium.
There are  two possible mechanisms than can provide a suppression:
\begin{itemize}
\item[(i)] The abundance of anomalous helium in the Galaxy may be significant, but in terrestrial matter
 a recombination mechanism could suppress this abundance below experimental upper limits \cite{Khlopov:2006dk,FKS}.
The existence of a new U(1) gauge symmetry, causing new Coulomb-like long range interactions between charged dark matter particles, is crucial for this mechanism. This leads inevitably to the existence of dark radiation in the form of hidden photons.
\item[(ii)] Free positively charged particles are already suppressed in the early Universe and the abundance
of anomalous helium in the Galaxy is negligible \cite{mpla,I}.
\end{itemize}
These two possibilities correspond to two different cosmological scenarios of dark atoms. The first one is
realized in the scenario with AC leptons, forming neutral AC atoms \cite{FKS}.
The second assumes a charge asymmetry  of the $O^{--}$ which forms the atom-like states with
primordial helium \cite{mpla,I}.

If new stable species belong to non-trivial representations of
the SU(2) electroweak group, sphaleron transitions at high temperatures
can provide the relation between baryon asymmetry and excess of
-2 charge stable species, as it was demonstrated in the case of WTC
\cite{KK,KK2,unesco,iwara}.

 After it is formed
in the Standard Big Bang Nucleosynthesis (BBN), $^4He$ screens the
$O^{--}$ charged particles in composite $(^4He^{++}O^{--})$ {\it
$OHe$} ``atoms'' \cite{I}.
In all the models of $OHe$, $O^{--}$ behaves either as a lepton or
as a specific ``heavy quark cluster" with strongly suppressed hadronic
interactions.
The cosmological scenario of the $OHe$ Universe involves only one parameter
of new physics $-$ the mass of O$^{--}$. Such a scenario is insensitive to the properties of $O^{--}$ (except for its mass), since the main features of the $OHe$ dark atoms are determined by their nuclear interacting helium shell. In terrestrial matter such dark matter species are slowed down and cannot cause significant nuclear recoil in the underground detectors, making them elusive in direct WIMP search experiments (where detection is based on nuclear recoil) such as CDMS, XENON100 and LUX. The positive results of DAMA and possibly CRESST and CoGeNT experiments (see \cite{DAMAtalk} for review and references) can find in this scenario a nontrivial explanation due to a low energy radiative capture of $OHe$ by intermediate mass nuclei~\cite{mpla,DMRev,DDMRev}. This explains the negative results of the XENON100 and LUX experiments. The rate of this capture is
proportional to the temperature: this leads to a suppression of this effect in cryogenic
detectors, such as CDMS. OHe collisions in the central part of the Galaxy lead to OHe
excitations, and de-excitations with pair production in E0 transitions can explain the
excess of the positron-annihilation line, observed by INTEGRAL in the galactic bulge \cite{DMRev,DDMRev,KK2,CKWahe}.

One should note that the nuclear physics of OHe is in the course of development and its basic element for a successful and self-consistent OHe dark matter scenario is related to the existence of a dipole Coulomb barrier, arising in the process of OHe-nucleus interaction and providing the dominance of elastic collisions of OHe with nuclei. This problem is the main open question of composite dark matter, which implies correct quantum mechanical solution \cite{CKW}. The lack of such a barrier and essential contribution of inelastic OHe-nucleus processes seem to lead to inevitable overproduction of anomalous isotopes \cite{CKW2}.

It has been shown \cite{KK,KK2,unesco,iwara,AHEP} that a two-component dark atom scenario is also possible and can be naturally realized in the framework of a WTC model, in which both stable double charged technilepton $\zeta^{--}$, playing the role of $O^{--}$, and positively double charged technibaryon $UU$ are predicted. Along with the dominant $\zeta^{--}$ abundance, a much smaller excess of positively doubly charged techniparticles $UU$ can be created. These positively charged particles are hidden in WIMP-like atoms, being bound to $\zeta^{--}$. In the framework of WTC such positively charged techniparticles can be metastable, with a dominant decay channel to a pair of positively charged leptons. We have shown in \cite{AHEP} that even a $10^{-6}$ fraction of such positively charged techniparticles with a mass of 1 TeV or less and a lifetime of~$10^{20} \s$,  decaying   to $e^+e^+$, $\mu^+ \mu^+$, and $\tau^+ \tau^+$  can explain the observed excess of cosmic ray positrons, being compatible with the observed gamma-ray background.

The anomalous excess of high-energy positrons in cosmic rays was first observed by PAMELA~\cite{PAMELA} and was later confirmed by AMS-02~\cite{AMS-2old}. These results generated widespread interest, since the corresponding effect cannot be explained by positrons of only secondary origin and requires primary positron sources, e.g. annihilations or decays of dark matter particles.
Recently AMS-02 collaboration has reported new results on positron and electron fluxes in cosmic rays~\cite{AMS2} and positron fraction~\cite{AMS1}. These measurements cover the energy ranges 0.5 to 700 GeV for electrons and 0.5 to 500 GeV for positrons and provides important information on the origins of primary positrons in the cosmic rays. In particular, new results show, for the first time, that above $\sim 200$ GeV the positron fraction no longer exhibits an increase with energy.
The possibility to explain the cosmic positron excess by the decays of $UU$ particles, comprising the tiny WIMP-component of dark matter in the considered scenario, was discussed in detail in~\cite{AHEP}. Here we estimate the optimal values of model parameters by achieving the best agreement with AMS-02 data on cosmic positron flux and FERMI-LAT data on diffuse gamma-ray flux ~\cite{FERMI}.\\

\section{Cosmic positron excess and fit to the latest AMS-02 data}\label{AMS}

In the considering scenario the metastable $UU$ particles, which together with $\zeta$ forms the subdominant component of dark matter, decays as $UU\rightarrow e^+e^+, \mu^+\mu^+, \tau^+\tau^+$ in principle with different branching ratios. All decay modes give directly or through cascades positrons and gamma photons.  The latter are hereafter referred to as final state radiation (FSR). The positron flux at the top of the Earth's atmosphere can be estimated as
\begin{equation}
F(E)=\frac{c}{4\pi}\frac{n_{\rm loc}}{\tau}\frac{1}{\beta E^2}
\int_E^{m/2} \frac{dN}{dE_0} Q(\lambda(E_0,E))dE_0,
\label{flux}
\end{equation}
where $n_{\rm loc}=\xi\cdot (0.3\,{\rm  GeV/cm^3})m_{UU}^{-1}$ is the local number density of $UU$ particles with $\xi=10^{-6}$, $dN/dE_0$ is the number of positrons produced in a single decay (obtained using Pythia 6.4~\cite{Pythia}), $\beta \sim 10^{-16}$ s$^{-1}$GeV$^{-1}$ and

\begin{equation}
Q=1-\frac{(\lambda-h)^2(2\lambda+4)}{2\lambda^3}\eta(\lambda-h)-\frac{2h(\lambda^2-r^2)}{3\lambda^3}\eta(\lambda-R),
\end{equation}

(see~\cite{AHEP} for details).\\

Below $\sim 10$ GeV behavior of positrons is affected by solar modulation. This effect can be in principle taken into account, using the force field model with two different parameters $\phi$ both for electrons and positrons, which can be easily adjusted in order to fit the data points at low energies. However, the effects of solar modulations are insignificant at the energies above $\sim 20$ GeV and thus for our analysis we consider the positron spectrum from 20 to 500 GeV. The positron background component was taken from ~\cite{secspec}.\\

Any scenario that provides positron excess is also constrained by other observational data, mainly from the data on cosmic antiprotons and gamma-radiation from our halo (diffuse gamma-background) and other galaxies and clusters. If dark matter does not produce antiprotons, then the diffuse gamma-ray background gives the most stringent and model-independent constraints.
For the FSR photons produced by $UU$ decays in our Galaxy, the flux arriving at the Earth is given by
\begin{equation}
F_{\rm FSR}^{\rm (G)}(E)=\frac{n_{\rm loc}}{\tau}\frac{1}{4\pi \Delta\Omega_{\rm obs}}\int_{\Delta\Omega_{\rm obs}}\frac{n(r)}{n_{\rm loc}}dld\Omega \cdot \frac{dN_{\gamma}}{dE},
\label{FSRgal}
\end{equation}
where we use an isothermal number density distribution  $\frac{n(r)}{n_{\rm loc}}=\frac{(5{\rm \,kpc})^2+(8.5{\rm \,kpc})^2}{(5{\rm \,kpc})^2+r^2}$, $r$ and $l$ are the distances from the Galactic center and the Earth respectively. We obtain the averaged flux over the solid angle $\Delta\Omega_{\rm obs}$ corresponding to $|b|>10^\circ$, $0<l<360^{\circ}$.\\
Out of our Galaxy, decays of $UU$ homogeneously distributed over the Universe should also contribute to the observed  gamma-ray flux. For FSR photons this contribution can be estimated as
\begin{eqnarray}
F_{\rm FSR}^{\rm (U)}(E)=\frac{c}{4\pi}\frac{\langle n_{\rm mod}\rangle}{\tau}\int \frac{dN}{dE}dt=
\frac{c\langle n_{\rm mod}\rangle}{4\pi\tau}\times\nonumber\\ \times \int_0^{\min(1100,\frac{m}{2E}+1)} \frac{dN}{dE_0}(E_0=E(z+1))\frac{H_{\rm mod}^{-1}dz}{\sqrt{\Omega_{\Lambda}+\Omega_m(z+1)^3}},
\label{FU}
\end{eqnarray}
where $z=1100$ corresponds to the recombination epoch, $\langle n_{\rm mod}\rangle$ is the current cosmological number density of $UU$, $H_{\rm mod}^{-1}=\frac{3}{2}t_{\rm mod}\sqrt{\Omega_{\Lambda}}\ln^{-1}\left(\frac{1+\sqrt{\Omega_{\Lambda}}}{\sqrt{\Omega_{m}}}\right)$ is the inverse value of the  Hubble parameter with $t_{\rm mod}$ being the  age of the Universe. 
$\Omega_{\Lambda}$ and $\Omega_m=1-\Omega_{\Lambda}$ are respectively the current vacuum and matter relative densities.

\begin{figure}
	\centering
	\includegraphics[scale=0.41]{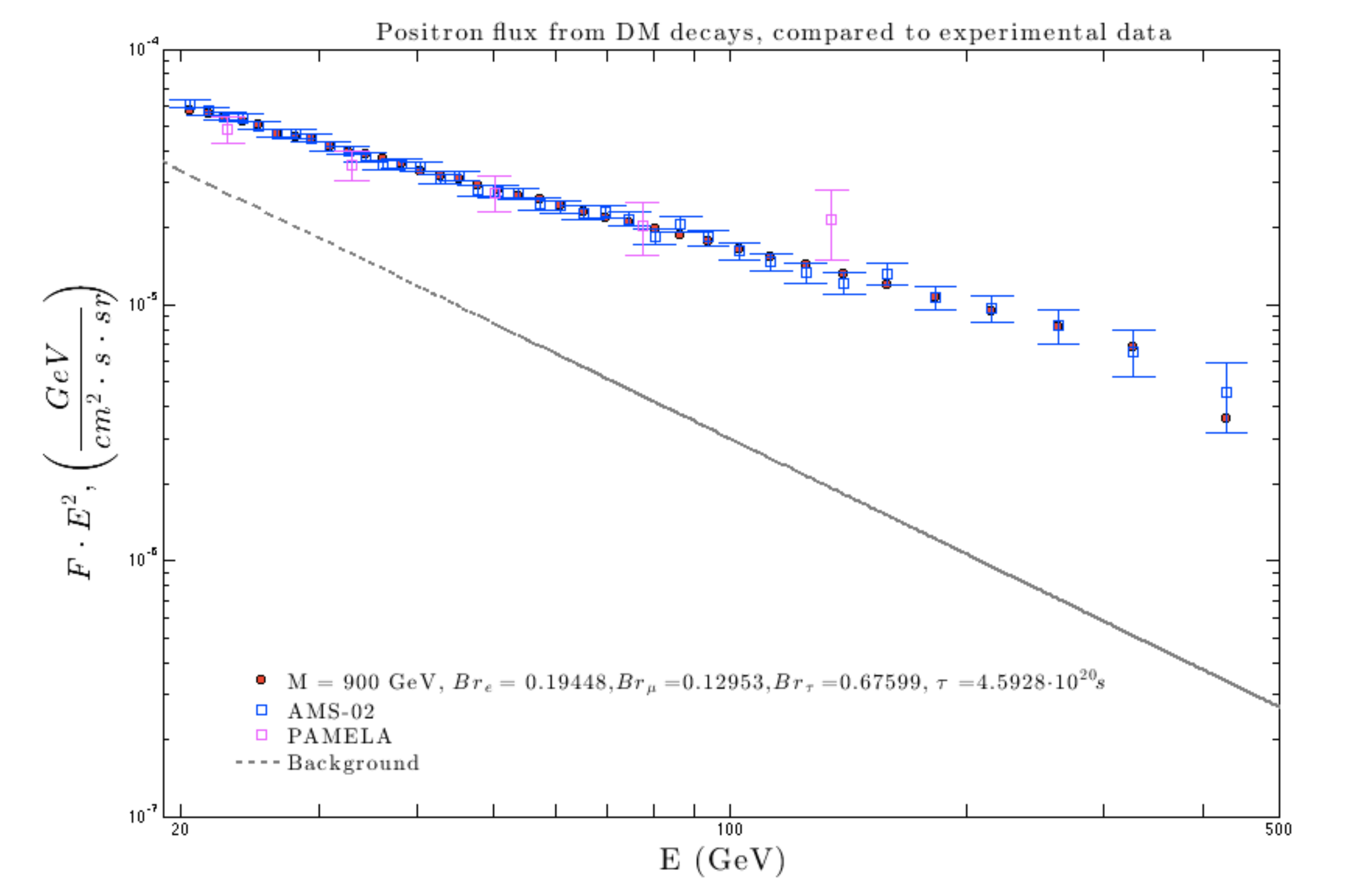}
	\caption{Positron flux from $UU$ decays compared to PAMELA and AMS-02 data}
	\label{posflux}
\end{figure}

\begin{figure}
	\centering
	\includegraphics[scale=0.43]{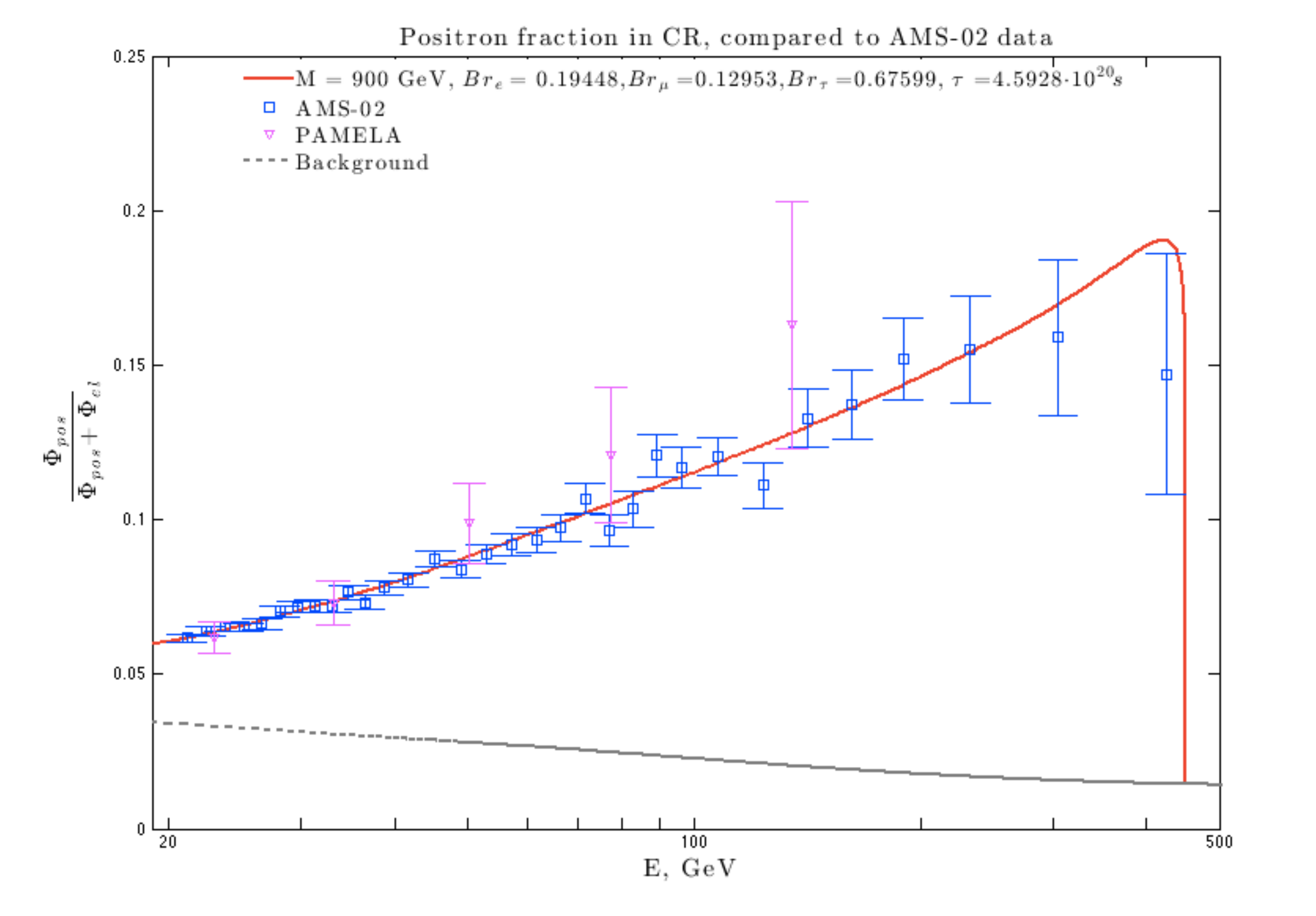}
	\caption{Positron excess due to $UU$ decays compared to PAMELA and AMS-02 data}
	\label{posfrac}
\end{figure}

\begin{figure}
	\centering
	\includegraphics[scale=0.45]{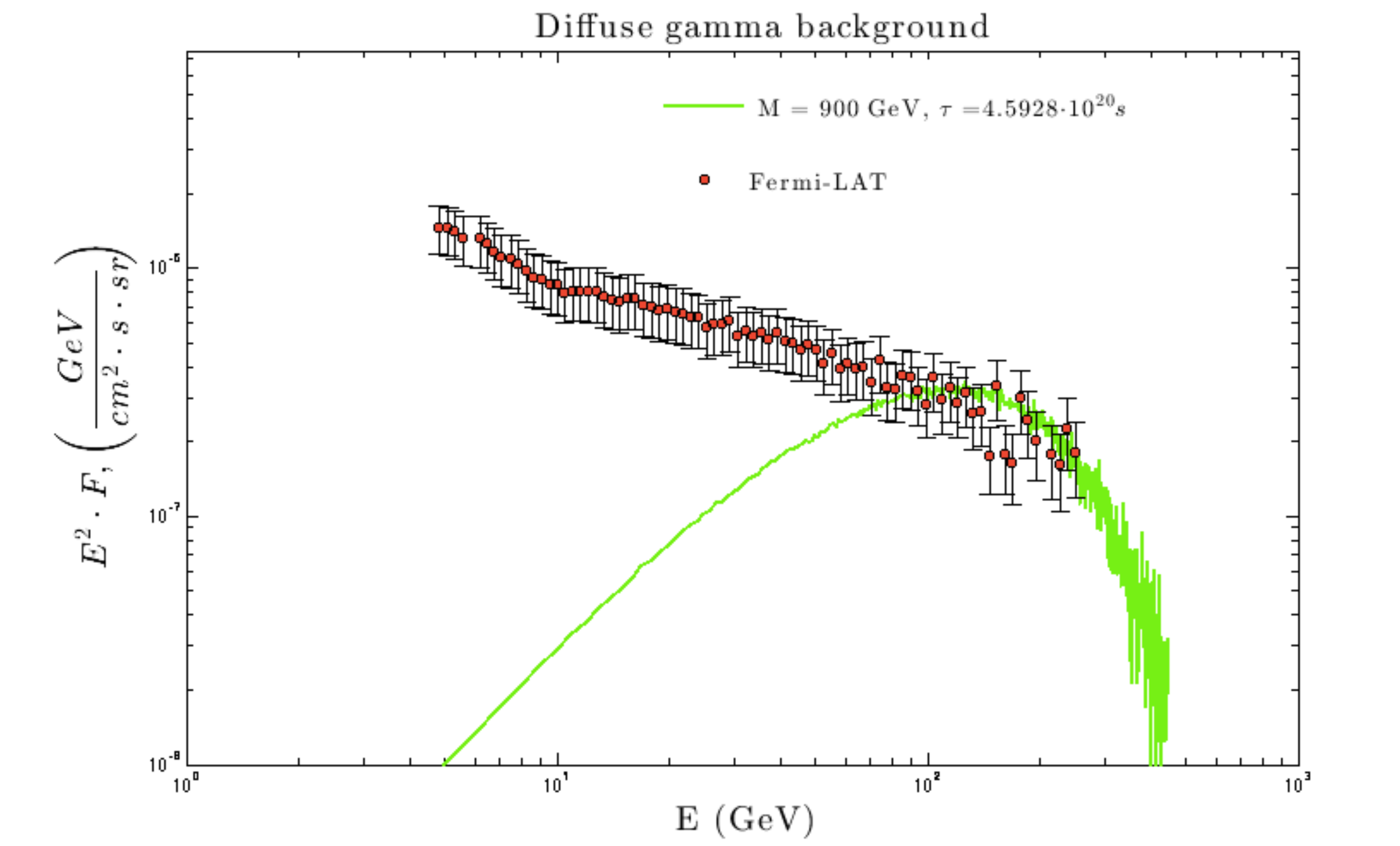}
	\caption{Gamma-ray flux from $UU$ decays in the Galaxy ($|b|\ge 10^{0}$) compared to the Fermi/LAT data on diffuse gamma-background}
	\label{gammabgr}
\end{figure}

Contribution into gamma-ray flux induced by scattering off background electromagnetic radiations of electrons and positrons from decays is small at high energy tail of spectrum, where observation data put the strongest constraint~\cite{AHEP}, and is not taken into account here.

The best-fit parameter values were obtained by fitting the curve, given by Eq.\eqref{flux}, to the AMS-02 data points in the least squares sense. Since the same parameters define the predicted gamma-ray flux in order not to contradict the FERMI-LAT data we extend the fitting procedure, involving several lowest points in the diffuse gamma-background spectrum above $\sim 100$ GeV to be fit.
For each choice of $m_{UU}$ from 700 to 1400 GeV we have evaluated the best-fit values of the lifetime $\tau$ and three branching ratios $Br_{e}$, $Br_{\mu}$ and $Br_{\tau} = 1 - Br_{e} - Br_{\mu}$. To choose the scenario, which is most consistent with the experimental data, a $\chi^2$ statistical test was used.
The best fit ($\chi^2 / n.d.f. = 0.57$) corresponds to the following parameter values: $M = 900 \GeV$, $\tau = 4.59 ~10^{20} \s$, $Br_{e} = 0.195$, $Br_{\mu} = 0.129$, $Br_{\tau} = 0.676$. Positron flux, positron fraction and gamma-ray flux in the best-fit case are shown in Figs. \ref{posflux},\ref{posfrac} and \ref{gammabgr} respectively.

\section{Conclusions}
Being the reflection of fundamental particle symmetry beyond the Standard model, the set of stable particles -- dark matter candidates -- can hardly be reduced to one single species~\cite{DMRev}. It makes natural to consider multi-component dark matter and one can hardly expect that various components put equal or comparable contribution into the total density. The situation with dominance of one component coexisting with some other subdominant components doesn't seem too exotic in this case.

Dark matter solution for the puzzles of dark matter searches can involve the form of neutral $OHe$ dark atoms made of stable heavy doubly charged particles and primordial He nuclei bound by ordinary Coulomb interactions. This scenario can be realized in the framework of Minimal Walking Technicolor, in which an exact relation between the dark matter density and baryon asymmetry can be naturally obtained. Strict conservation of technilepton charge together with approximate conservation of technibaryon charge results in the prediction of two types of doubly charged species with strongly unequal excess -- dominant negatively charged technileptons $\zeta^{--}$ and a strongly subdominant component of technibaryons $UU^{++}$, bound with $\zeta^{--}$ in
  a sparse component of WIMP-like dark atoms ($\zeta^{--}UU^{++}$). Direct searches for WIMPs put severe constraints on the presence of this component. However we have demonstrated in~\cite{AHEP}  that
 the existence of a metastable positively doubly charged techniparticle, forming this tiny subdominant WIMP-like dark atom component and satisfying the direct WIMP searches constraints, can play an important role in the indirect effects of dark matter. We found that decays of such positively charged constituents of WIMP-like dark atoms to the leptons $e^+e^+, \mu^+\mu^+,\tau^+ \tau^+$ can explain the observed excess of high energy cosmic ray positrons, while being compatible with the observed gamma-ray background. These decays are naturally  facilitated by GUT scale interactions. The best fit of the data takes place for a mass of this doubly charged particle of 1 TeV or below making it accessible in the next run of LHC. Our refined analysis of the best fit description of the recent data of the AMS-02 experiment, presented here, can provide a crucial test for the decaying dark atom hypothesis in the experimental searches for stable doubly charged lepton-like particles at the LHC.
\section*{Acknowledgements}
We thank Chris Kouvaris for collaboration in development of the presented approach. The work by M.Kh. and M.L. on initial cosmological conditions was supported by the Ministry of Education and Science of Russian Federation, project 3.472.2014/K  and the work of K.B., M.Kh. and M.L. on the forms of dark matter was supported by grant RFBR 14-22-03048.


\end{document}